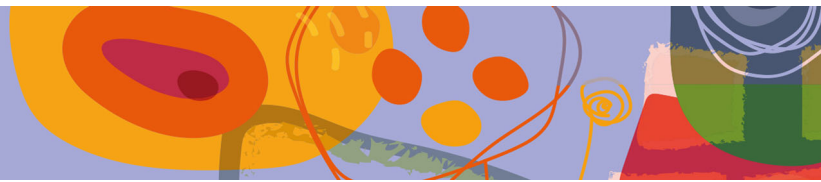

## ARTICLE

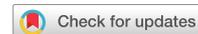

# Artificial intelligence and the Gulf Cooperation Council workforce: adapting to the future of work

Mohammad Rashed Albous[1✉], Melodena Stephens[2] & Odeh Rashed Al-Jayyousi[3]

The rapid expansion of artificial intelligence (AI) in the Gulf Cooperation Council (GCC) raises a central question: are investments in compute infrastructure matched by an equally robust build-out of skills, incentives, and governance? Grounded in socio-technical systems (STS) theory, this mixed-methods study audits workforce preparedness across Kingdom of Saudi Arabia (KSA), the United Arab Emirates (UAE), Qatar, Kuwait, Bahrain, and Oman. We combine term frequency–inverse document frequency (TF–IDF) analysis of six national AI strategies (NASs), an inventory of 47 publicly disclosed AI initiatives (January 2017–April 2025), paired case studies, the Mohamed bin Zayed University of Artificial Intelligence (MBZUAI) and the Saudi Data & Artificial Intelligence Authority (SDAIA) Academy, and a scenario matrix linking oil-revenue slack (technical capacity) to regulatory coherence (social alignment). Across the corpus, 34/47 initiatives (0.72; 95% Wilson CI 0.58–0.83) exhibit joint social–technical design; country-level indices span 0.57–0.90 (small n; intervals overlap). Scenario results suggest that, under our modeled conditions, regulatory convergence plausibly binds outcomes more than fiscal capacity: fragmented rules can offset high oil revenues, while harmonized standards help preserve progress under austerity. We also identify an emerging two-track talent system, research elites versus rapidly trained practitioners, that risks labor-market bifurcation without bridging mechanisms. By extending STS inquiry to oil-rich, state-led economies, the study refines theory and sets a research agenda focused on longitudinal coupling metrics, ethnographies of coordination, and outcome-based performance indicators.

[1] Abdullah Al Salem University, Khaldiya, Kuwait. [2] Mohammed Bin Rashid School of Government, Dubai, United Arab Emirates. [3] Arabian Gulf University, Manama, Bahrain. ✉email: Mohammad.albous@aasu.edu.kw

## Introduction

I f we step back and view today's advances through the long lens of general-purpose technologies, a familiar pattern emerges. Artificial intelligence (AI) is reshaping cognitive work at scale, raising productivity while disrupting established job structures. Machine-learning systems, large language models (LLMs), and autonomous agents promise breakthroughs that the International Monetary Fund (2024) estimates could reshape or replace up to 40% of jobs worldwide, while the World Economic Forum's Future of Jobs Report (2025) foresees simultaneous waves of job creation, destruction, and wholesale skill recombination (Bonfiglioli et al., 2025). Yet the pressures that spur digital abundance also widen the gap between workers who can complement intelligent machines and those who cannot. Macro-level studies that warn of "job polarization" (Acemoglu & Restrepo, 2020, 2022) and the vulnerability of professions once thought safe (Frey & Osborne, 2017, 2023) reinforce a central lesson: technology's impact is not pre-ordained but is mediated by human choices, complementary investments, and enlightened governance (see also Autor et al., 2024; Albanesi et al., 2025).

Yet aggregate averages can obscure pronounced cross-regional variation. Most of the empirical scaffolding behind these insights, however, rests on data from North America, Europe, and a handful of other advanced industrial economies. We therefore know far less about how regions with distinctive labor institutions and resource endowments will navigate similar technological headwinds (Csernatoni, 2024; Roche et al., 2023). Nowhere is this more evident than in the Gulf Cooperation Council (GCC) – Kingdom of Saudi Arabia (KSA), the United Arab Emirates (UAE), Qatar, Kuwait, Bahrain, and Oman – which illustrates this blind spot vividly. Buoyed by hydrocarbon wealth and extensive public-sector employment, GCC states have set ambitious visions to pivot toward knowledge-intensive growth and to claim leadership in AI. Whether those ambitions translate into broadbased prosperity depends on how effectively each country mobilizes policy, redesigns education and training, and builds institutions that allow national and expatriate workers to thrive alongside smart machines.

To explain why context matters so much, we need a framework that marries the technical with the social. Socio-technical Systems Theory (STS), first articulated by Emery and Trist (1960) and refined by Cherns (1976), offers a powerful lens for analyzing this challenge. STS argues that durable performance gains arise only when technical subsystems, algorithms, data architectures, high-performance computing (HPC), and social subsystems, skills, incentives, governance, culture, are jointly optimized. Technological interventions that ignore the social lattice tend to underperform, just as workforce programs detached from technical realities rarely scale. The GCC's combination of oil-funded infrastructure, expatriate labor regimes, and state-led development models makes the region a natural laboratory for testing whether socio-technical alignment can be achieved under conditions that differ markedly from those studied in Western contexts.

Against this backdrop, the present research asks: How well are GCC nations, through their policies, training programs, and governance frameworks, preparing their workforces for an AIpowered future of work? Addressing this question not only fills a geographic gap in the literature but also enriches theoretical debates on how state capacity, resource dependence, and institutional design shape technology's labor-market effects.

To answer, the paper weaves together four complementary strands of evidence within a unified STS frame. It first reviews every publicly available National AI Strategy (NAS) in the GCC, tracing stated goals, funding commitments, and accountability mechanisms. It then maps 47 publicly disclosed AI initiatives, from smart-city pilots to start-up accelerators, to gauge implementation momentum on the ground. Comparative case studies juxtapose an elite graduate-level AI university, the Mohamed bin Zayed University of Artificial Intelligence (MBZUAI), from the UAE, with a large-scale vocational up-skilling academy, the Saudi Data & Artificial Intelligence Authority (SDAIA) Academy from the KSA, revealing trade-offs between depth and breadth of talent formation. Finally, a scenario analysis explores alternative futures under varying oil-revenue trajectories and degrees of regulatory coordination. Together, these perspectives yield a granular, yet holistic portrait of how a resource-rich, state-led region is grappling with the promises and perils of the AI revolution.

The paper proceeds as follows: Literature review situates the study within global debates on AI, labor, and STS; Methodology outlines the four-phase, mixed-methods approach that links policy texts, initiative mapping, case studies, and scenario modeling; Findings reports the main evidence on GCC strategies, skills alignment, governance gaps, and alternative futures; Discussion interprets these results, highlighting where sociotechnical balance is emerging and where it still falters, and proposes policy and research implications; Conclusion distills the overall contribution and notes key limitations and next steps.

## Literature review

Scholarship on AI and work converges around three pillars: macroeconomic projections, theories of automation-driven job and task disruption, and real-time labor-market evidence from platforms such as LinkedIn and Microsoft. This literature review synthesizes those strands in turn and closes with their implications for the GCC's rapidly evolving workforce.

**Macroeconomic view**. A global macroeconomic perspective is provided by two key IMF publications that shed light on AI's present and future impact. First, the IMF's Staff Discussion Note, Gen-AI: Artificial Intelligence and the Future of Work (International Monetary Fund, 2024), explores how AI may reshape employment structures across advanced, emerging, and developing economies. According to this note, approximately 40% of global employment is susceptible to AI integration. In advanced economies, the figure rises to around 60%, reflecting the prevalence of cognitive-task-oriented roles, while it stands at 40% for emerging markets and 26% for low-income countries. These variations suggest that wealthier, more technologically developed regions may experience AI's impact more rapidly, both in terms of opportunities and displacements.

The Staff Discussion Note also highlights key demographic nuances: women and college-educated workers are more likely to be in positions exposed to AI but can also more readily capitalize on AI's benefits. By contrast, older workers may face difficulties adapting, potentially exacerbating age-based inequalities. Income and wealth inequalities could likewise intensify if AI disproportionately complements high-income tasks, raising concerns over labor income distribution and capital returns. Nevertheless, the IMF emphasizes that productivity gains from AI could lift overall income levels, provided governments adopt proactive policies such as investments in digital infrastructure, skills training, and inclusive regulatory frameworks. Advanced economies are urged to update regulations that facilitate AI adoption, while emerging and developing economies must reinforce their digital infrastructures to avoid widening global divides. Within the GCC, baseline digital capacity also differs, as tracked by the International Telecommunication Union ICT Development Index (ITU, 2024).

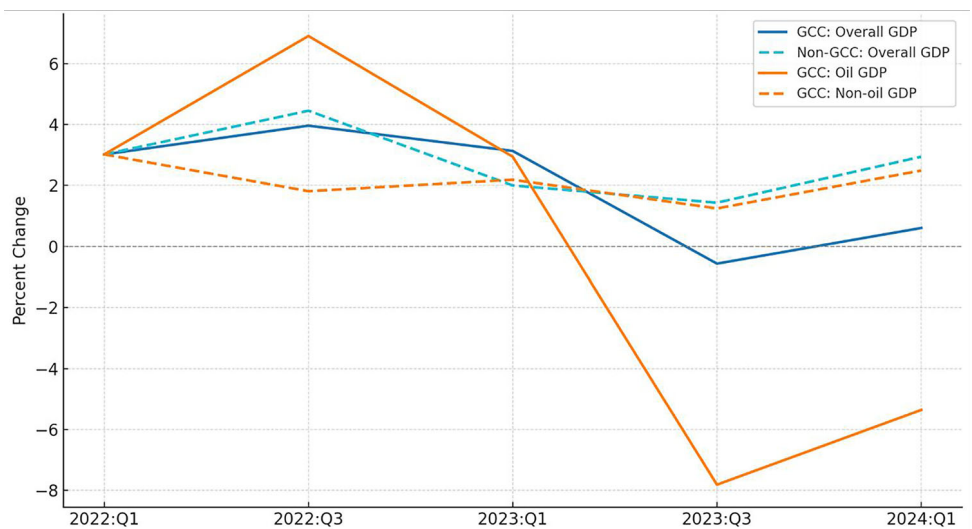

**Fig. 1 MENA oil exporters: real GDP growth, 2022–2024.** Quarterly oil vs. non-oil real GDP growth for GCC oil exporters; authors' visualization from IMF Regional Economic Outlook (Oct 2024). *Source:* IMF REO ME&CA 2024; authors' calculations.

In addition to these AI-focused findings, the IMF's Regional Economic Outlook for the Middle East and Central Asia (International Monetary Fund, 2024, October) offers critical insights into the broader macroeconomic environment in which AI adoption will occur. The report underscores the region's need to navigate an evolving geoeconomic landscape, shaped by ongoing geopolitical shifts, fluctuating oil revenues, and varying levels of fiscal space across GCC countries.

Figure 1, based on IMF data, highlights the contrasting growth trajectories of the GCC's oil and non-oil sectors from 2022 through early 2024. While non-oil GDP has remained relatively stable, oil GDP has experienced sharp declines, particularly in late 2023, following extended OPEC+ production cuts. As a result, the overall GDP trend underscores how the resilience of non-oil sectors can partially offset the volatility of oil revenues. This divergence has important implications for AI readiness: countries with more stable and diversified fiscal bases, bolstered by consistent non-oil growth, are better positioned to invest in long-term AI infrastructure and human capital. At the same time, the GCC's position in a conflict-prone region adds another layer of geopolitical uncertainty to fiscal planning and technology-investment horizons, a dynamic that became especially evident in mid-2025.

By emphasizing the need for sound fiscal management and structural reforms, the Regional Economic Outlook suggests that GCC governments capable of maintaining stable budgets and targeted public investments will be better positioned to scale up AI readiness programs, particularly in infrastructure and workforce development, than those constrained by external economic shocks. These broader macroeconomic drivers form the foundation upon which AI-specific policies, such as reskilling initiatives, governance frameworks, and sectoral AI roadmaps, must be built to ensure a sustainable transition to AI-driven industries.

*WEF's future of jobs report 2025.* Building upon the IMF's macro-level observations, the World Economic Forum's Future of Jobs Report 2025 presents in-depth labor market forecasts shaped by technological advancements, demographic change, and economic realignment. Drawing on data from over 1,000 employers across 55 economies, the report projects that by 2030, approximately 170 million jobs will be created, while 92 million existing jobs will be displaced, resulting in a net employment gain of 78 million roles. This reflects a structural churn involving 22% of the current global workforce, as projected in the WEF report (World Economic Forum, 2025).

These projections are visualized in Fig. 2, which illustrates the expected distribution of jobs created, destroyed, and remaining stable through the end of the decade. The majority of positions (~78%) are expected to remain stable, though uncertainty exists, while job creation outpaces displacement in aggregate. This figure is based on data synthesized from the World Economic Forum's Future of Jobs Survey 2024 and employment estimates from the International Labor Organization's ILOSTAT database.

The report also highlights that 39% of core workforce skills are expected to change significantly by 2030, with surging demand for roles such as AI and Machine Learning Specialists, Data Analysts, Renewable Energy Engineers, and Cybersecurity Experts. Meanwhile, clerical and mid-level administrative positions face heightened risk of automation and obsolescence. These shifts emphasize the urgent need for large-scale reskilling and adaptive education systems across all sectors.

Crucially, the trajectory of specific occupations remains hard to forecast with precision: roles once hyped, such as "prompt engineer," have already lost salience. Some industry leaders anticipate substantial automation of software-development tasks; however, precise shares remain uncertain. As of 2025, public statements indicated that a material share of code within large firms is already AI-assisted, but forward-looking percentages are speculative, Microsoft's CEO suggested up to 30% of code was AI-assisted (TechCrunch, 2025). This inherent unpredictability in job-market evolution is itself an emerging source of risk.

*Regional empirical evidence from the GCC.* Recent evidence indicates that adoption dynamics in the GCC differ markedly from those in North America and Europe. A 2024 survey of 140 C-suite leaders across eight GCC industries, conducted by McKinsey and the GCC Board Directors Institute, found that 73% of organizations had piloted generative-AI applications, yet only 11% had realized measurable value, citing talent shortages and data-governance frictions as the main blockers. The projected economic upside ranges from US $21 billion to US $35 billion, equivalent to as much as 2.8% of current non-oil GDP (Isherwood et al., 2024).

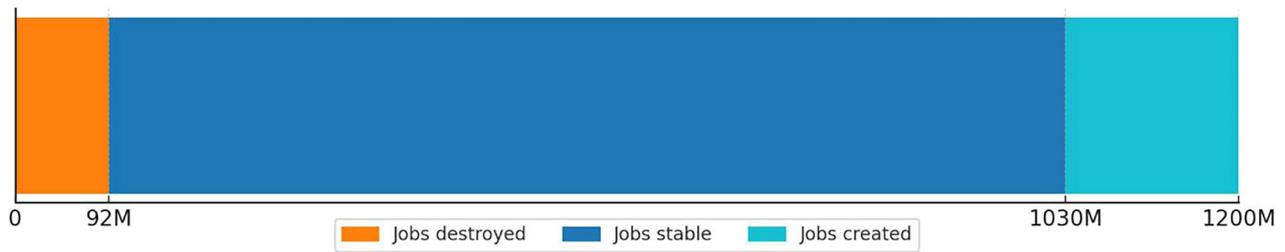

**Fig. 2 Projected global employment change by 2030.** Estimated jobs created, displaced, and stable through 2030 based on WEF Future of Jobs Survey 2024 and ILOSTAT; authors' redraw. WEF (2025) report; ILOSTAT.

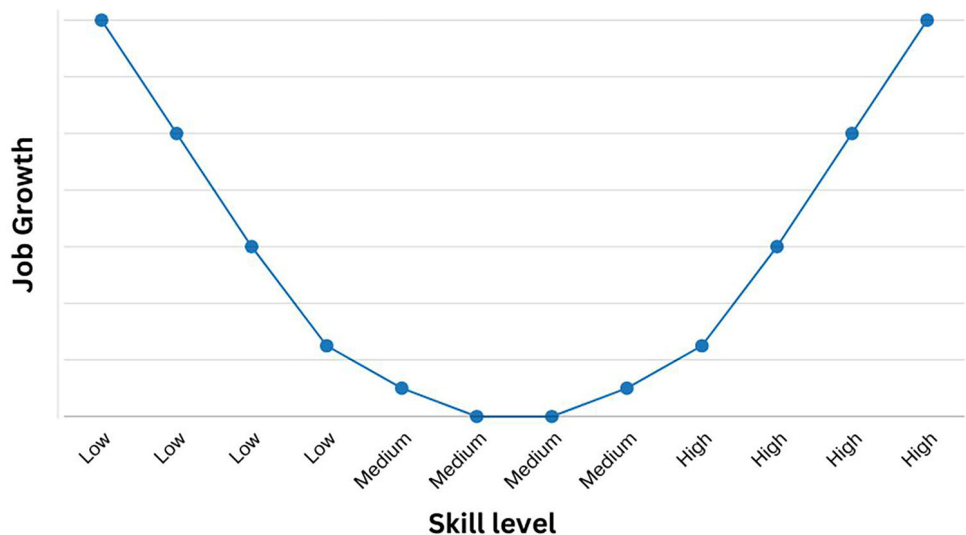

**Fig. 3 Job growth versus skill level (illustrative U-shape).** Conceptual schematic of polarization dynamics based on Acemoglu & Restrepo. Authors' illustration based on cited literature.

Worker-level data reinforce this leadership-interest gap. In a 19-country poll of 25,000 employees, 74% of UAE and 68% of KSA respondents reported weekly Gen-AI use, but 61% said their employers provide insufficient training, the highest anxiety score of any high-income region (Studer, 2024).

Governments are responding with large-scale capacity-building. KSA's SDAIA reports 779,000 citizens trained in data- and AI-skills as of May 2025, including 9,775 specialists and 260 scientists (Malin, 2025). In Qatar, a labor-market audit estimates that 46.5% of domestic tasks could be AI-augmented without triggering wage polarization, owing to the country's high tertiary-education share (Chauhan, 2020).

### Theoretical perspectives: job polarization and task-based automation

*Job polarization: Acemoglu & Restrepo.* Acemoglu and Restrepo (2020, 2022) advance a framework of job polarization, observing that routine-intensive, middle-skill positions, often found in administrative, clerical, or production roles, are particularly susceptible to automation. These dynamics foster a U-shaped employment distribution, in which the highest- and lowest-skill jobs expand while the mid-tier shrinks. Figure 3 visually illustrates this U-shaped pattern, highlighting the relative decline of middle-skill employment over time. If unaddressed, job polarization can exacerbate income inequality and destabilize labor markets by pushing displaced workers into lower-paying jobs or unemployment, see also Acemoglu and Restrepo (2024) for updated implications for wages and inequality.

Although such polarization primarily affects advanced economies, Acemoglu and Restrepo (2023) note that emerging economies with large public-sector workforces can face similar pressures. The challenge lies in enabling mid-level workers to transition into roles requiring complex cognitive tasks or specialized technical skills, areas where AI is less likely to entirely replace human labor.

*Susceptibility to automation: Frey & Osborne.* Frey and Osborne (2017, 2023) introduce a task-based analysis that broadens the focus from purely manual labor to include cognitive tasks in areas such as finance and customer service. As AI continues to advance, a growing share of white-collar and knowledge-based occupations faces potential automation. While culturally and practically resistant fields like healthcare and education may remain partially insulated, the trajectory of AI and robotics development suggests that even these sectors are not entirely exempt. Figure 4 illustrates this trend, showing how tasks that are both routine and manual exhibit the highest susceptibility to automation, whereas non-routine cognitive tasks remain more resistant. This underscores the importance for governments and organizations to analyze the specific task composition of roles in order to better forecast potential disruption and target reskilling efforts effectively and set up guardrails to protect valuable skills that may face obsolescence.

### Empirical insights from LinkedIn and Microsoft

*LinkedIn's Work Change Report: AI is coming to work.* Drawing on data from over one billion professionals and 69 million companies, LinkedIn's Work Change Report (2025), AI Is Coming to Work, illustrates how AI is fundamentally reshaping the global labor landscape. The report projects that by 2030, 70% of the skill sets required in most occupations will have undergone significant transformation. New workforce entrants are expected to change roles at nearly twice the rate of their counterparts from 15 years ago, reflecting a heightened demand for adaptability in an AI-augmented economy.

Among the most striking indicators of this shift is the global surge in AI-related hiring. Over the past 8 years, AI hiring has grown by >300%. This trend is visualized in Fig. 5, which charts the pace of AI hiring relative to overall hiring between February 2022 and August 2024. The sharp upward trajectory, particularly post-2023, underscores the acceleration of AI adoption across industries and geographies.

Yet, despite this hiring boom, there is a persistent disconnect between leadership expectations and hiring criteria: only one in 500 current job postings explicitly lists AI proficiency as a requirement. This suggests that while organizations increasingly value AI capabilities, formal HR frameworks have yet to fully integrate these competencies.

Regionally, the UAE stands out, registering an 80-fold increase in the number of LinkedIn members adding AI skills to their profiles since 2016, far exceeding global averages. This rapid uptake reflects strong policy momentum around AI but also signals the risk of a growing skills mismatch if educational systems and training programs fail to keep pace with evolving industry demands.

*Microsoft's 2024 Work Trend Index Annual Report*. In the 2024 Work Trend Index Annual Report, co-published by Microsoft and LinkedIn, surveys of 31,000 workers across 31 countries reveal the rapid mainstreaming of AI in the workplace. A striking 75% of knowledge workers already use AI tools, and 78% bring their own AI solutions (BYOAI) into environments that often lack formal governance policies. While 79% of business leaders recognize AI as critical for staying competitive, 60% admit their organizations still lack a clear strategy for implementation, as reported in the 2024 Work Trend Index (Microsoft & LinkedIn, 2024).

One of the report's most notable insights is the emergence of the "AI Power User," a distinct group of employees who integrate AI into their workflow multiple times per week and report saving over 30 min daily. These users are not only more productive but also reimagine task flows, boost creativity, and feel more motivated in their roles.

As illustrated in Fig. 6, AI power users self-report significantly stronger positive outcomes compared to other user types, such as explorers, novices, and skeptics. The figure highlights clear differences across dimensions like task focus, creativity, and motivation, demonstrating how frequent and strategic AI usage correlates with improved work experience and performance. This segmentation underscores the transformative potential of empowering a broader share of the workforce to become AI-literate and strategically fluent, but the report does not highlight how these productivity gains lead to a robust knowledge-sharing economy.

**A Socio-Technical Systems (STS) perspective on AI and work**. The mixed evidence above raises a fundamental question: why do ostensibly similar AI deployments produce sharply different labor-market outcomes across countries, sectors, and even adjacent organizations? STS, first articulated in the Tavistock coalmining studies (Emery & Trist, 1960) and later codified by Cherns (1976) and Pasmore (1988), offers a durable explanatory framework. STS argues that work outcomes depend on the joint optimization of two interdependent subsystems: the technical (algorithms, data pipelines, hardware) and the social (skills, incentives, organizational structures, regulatory rules). Because of this interdependence, the principle of equifinality holds, similar technologies can yield divergent results if embedded in different social contexts. Conversely, well-aligned social arrangements can compensate for less advanced technology (Baxter & Sommerville, 2011).

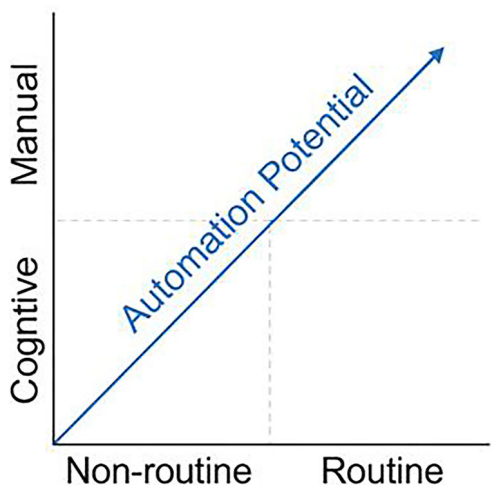

**Fig. 4 Task susceptibility to automation.** Conceptual matrix contrasting routine/non-routine and manual/cognitive tasks, synthesized from Frey & Osborne. Authors' illustration based on cited literature.

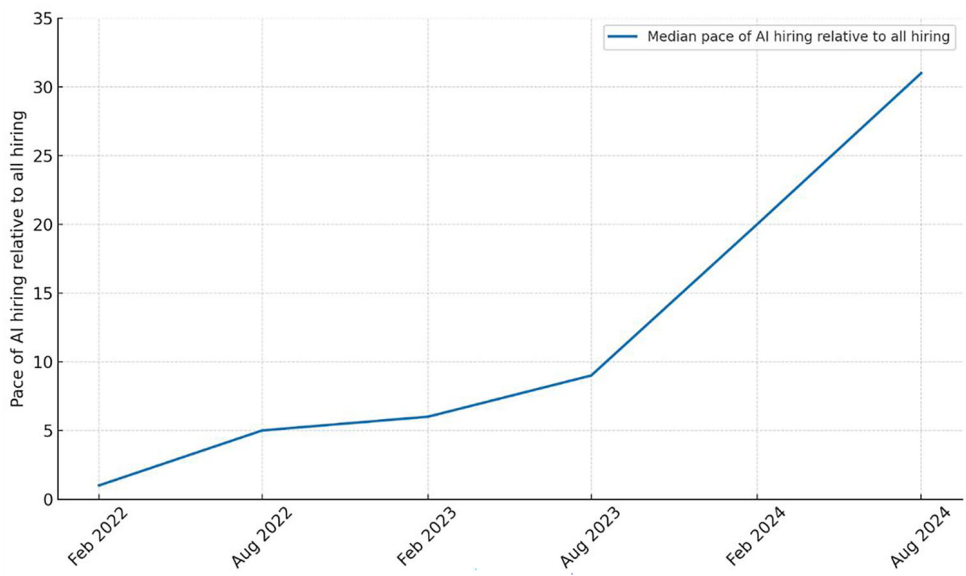

**Fig. 5 Change in global AI hiring relative to all hiring, 2022–2024.** Relative hiring index; authors' redraw from LinkedIn Work Change Report series. LinkedIn Economic Graph (2025).

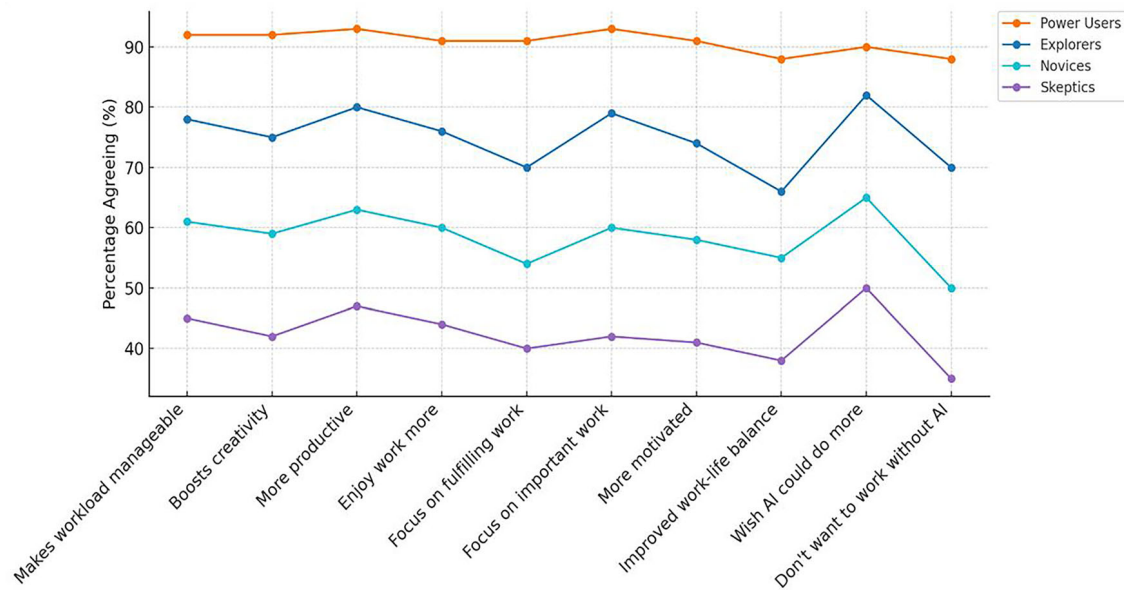

**Fig. 6 Self-reported outcomes for AI user types.** Differences across focus, creativity, and motivation; authors' redraw from Microsoft–LinkedIn Work Trend Index (2024). 2024 WTI.

Recent scholarship has applied STS to contemporary digital transformations, underscoring its relevance for AI. Thomas (2024) shows how knowledge-management practices mediate AI's productivity effects in European service firms, while Kudina and van de Poel (2024) develop an STS framework to trace value-laden trade-offs in algorithmic decision-making. Johnson and Verdicchio (2024) extend the analysis to the ethical domain, demonstrating how AI and human values become co-constituted in socio-technical "entanglements". Supply-chain scholars likewise foreground STS in RFID-driven automation (Zhang et al., 2025), revealing that governance routines, not sensor accuracy alone, determine downstream labor impacts.

Yet the empirical base of AI-STS research remains overwhelmingly Western and focused on advanced industrial economies, leaving oil-dependent, state-led labor markets under-examined. The GCC presents a distinctive test bed: abundant capital facilitates rapid technical upgrades, but social subsystems, public-sector career norms, expatriate labor regimes, nascent data-governance laws, may lag. Integrating an STS lens therefore bridges the macro statistics reviewed earlier (IMF, WEF) with the micro realities of work redesign, offering a structured way to diagnose whether GCC AI initiatives achieve genuine socio-technical alignment or merely enact "technology push".

**Implications for the GCC.** Despite the rapidly expanding scholarship on AI, most of the empirical evidence still comes from North America, Western Europe and a handful of other advanced industrial economies. Because datasets, case studies and institutional narratives remain so heavily Western, we have only a partial understanding of how economies that differ sharply in labor structures, policy traditions and resource endowments, such as the GCC states, will confront the same technological headwinds (Csernatoni, 2024; Roche et al., 2023).

Generative AI systems, LLMs and advanced HPC resources magnify both opportunities and risks for the GCC. Policymakers, guided by diversification agendas, view AI as a growth engine but also recognize its disruptive potential, especially for the mid-skill public-sector administrative roles that dominate local employment structures and are highly automatable.

On the upside, an expanding portfolio of in-country training schemes, ranging from SDAIA's short-cycle academies to MBZUAI's doctoral fellowships, signals a sustained commitment to building specialized AI capability (MBZUAI, 2024, 2025; SDAIA, 2020, 2023). Parallel investment in hyperscale HPC clusters, sovereign cloud regions and Arabic-first LLMs underscores a determination to move up the digital value chain. The payoff is already visible in cross-border talent flows: the Stanford AI Index (2025) ranks the UAE third and KSA eighth worldwide for net AI-talent migration, confirming that the GCC is now a preferred destination for globally mobile specialists (Maslej et al., 2025). Yet success on that front exposes two structural vulnerabilities. First, as Malit and Al-Shanqityi (2025) warn, large-scale recruitment of high-skilled expatriates can amplify the "brain-drain/brain-gain" dynamic, widening capability gaps with lower-income source countries and entrenching domestic dependence on imported expertise. Second, and more critically for innovation policy, not a single GCC hub appears in 2024 WIPO's list of the world's top-100 Science, Technology and Innovation clusters (World Intellectual Property Organization, 2024). The absence suggests that headline talent inflows and hardware buildouts have not yet crystallized into dense, self-sustaining innovation ecosystems, making ecosystem coherence, rather than raw capacity, the region's next strategic frontier.

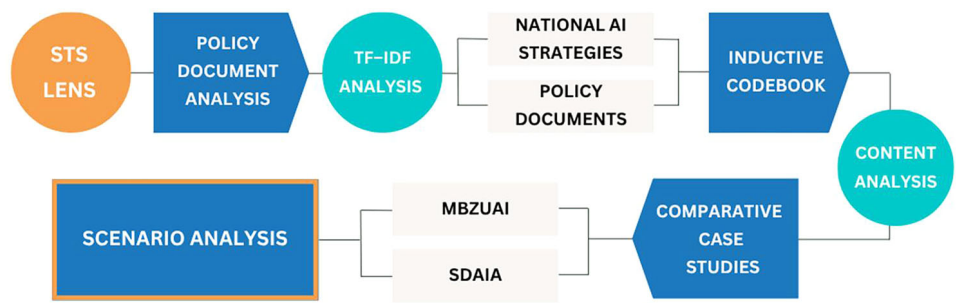

**Fig. 7 Four-phase research design with STS integration.** Study workflow linking policy analysis, inductive coding, case studies, and scenario modeling. Authors' original figure.

Grass-roots adoption of AI tools by individual employees further complicates the picture. While bottom-up experimentation accelerates learning, it can easily outrun governance, raising concerns about data privacy, security and ethical oversight, a pattern also noted by Studer (2024). Moreover, heavy reliance on imported talent carries its own risks. Without robust local education pipelines, from basic digital literacy through advanced algorithmic design, the region could lock itself into permanent external dependence. Finally, regulators have thus far leaned on voluntary guidelines and policy charters rather than binding statutes, reflecting a "soft-law" approach to AI governance that still leaves significant accountability gaps (Albous et al., 2025).

Over the long run, therefore, sustainable AI ecosystems in the GCC will require co-evolution of three elements: (1) enforceable governance that keeps pace with rapid experimentation; (2) inclusive training pathways that move citizens into higher-value roles rather than simply layering expatriate expertise on top of existing structures; and (3) talent-mobility regimes that convert today's inbound migration advantage into durable knowledge transfer rather than transient staffing relief. Absent such coordinated development, AI adoption could deepen job polarization, creating elite tech roles while displacing mid-level administrative positions, and ultimately undermine the region's quest for inclusive, knowledge-based growth.

### Methodology

This study adopts a four-phase, mixed-methods design, policy analysis, inductive content coding, comparative case studies, and scenario modeling, to evaluate how GCC countries are preparing their workforces for an AI-driven future of work. STS Theory provides the common thread linking the phases: each inquiry asks whether organizational initiatives jointly optimize the technical subsystem (algorithms, data, HPC, automation tools) and the social subsystem (skills, incentives, governance, culture) within the region's distinctive environmental context of oil economics and expatriate labor markets. Figure 7 visualizes the research design and shows, for each phase, the STS concepts operationalized.

### Phase 1: Policy document analysis (macro "design premise" level).

The first phase examines the six NASs issued by GCC member states, together with key policy reports and eco-system white papers, to clarify each country's objectives, implementation mechanisms and stakeholder roles in AI-driven workforce development. Building on a Term Frequency × Inverse Document Frequency (TF–IDF) procedure (Manning, 2009; Robertson, 1977), we quantified the salience and distinctiveness of terms appearing in titles, mandates and implementation sections (Lewis et al., 2020). The TF–IDF computation is presented in Equation (1):

$$TF - IDF(t, d) = TF(t, d) \times IDF(t)$$

where $TF(t, d)$ is the raw frequency of term $t$ in document $d$, and $IDF(t)$ captures the inverse prevalence of $t$ across the corpus

All NASs PDFs were converted to machine-readable text with *pdfplumber*; pages containing only images underwent optical-character-recognition via *Adobe* Optical Character Recognition (OCR) and *Tesseract*, affecting roughly 17% of the total corpus. Post OCR validation kept the error rate below 5%, ensuring reliable quantitative and qualitative interrogation. After the TF–IDF scores were generated, every extracted term was manually mapped to one of three STS categories:

- **Technical**. References to HPC, data centres, NLP, LLMs, cloud architecture, cybersecurity tools.
- **Social**. References to upskilling/ reskilling, organizational redesign, labor law, ethics committees, incentive schemes.
- **Environment**. References to oil-revenue assumptions, expatriate labor policies, demographic priorities, Arabic-language AI.

This ex-post coding links the lexical emphasis of each NAS to the "design premises" of STS (Emery & Trist, 1960; Cherns, 1976): a strategy is considered balanced when technical and social terms appear in tandem and are contextualised by environmental constraints. Country level bar chart (shown in the Codebook) visualize the relative weight of each category, providing an initial gauge of joint optimization balance across the six GCC strategies.

Finally, a structured qualitative review contextualised the quantitative scores, tracing how specific term clusters, for example ("LLM + Arabic NLP" or "skills + Vision 2030") translate into concrete governance instruments and funding pipelines. The resulting dataset serves as the macro-level foundation for the inductive content analysis and case-study work that follow.

### Phase 2: Inductive codebook development (meso, program-implementation level).

A corpus of 47 publicly disclosed AI initiatives (January 2017– April 2025) was coded using a three-category STS frame (Technical, Social, Environment), with the latter coded opportunistically when texts make context explicit:

| Layer | Code family (illustrative labels not exhaustive) | Coding rule |
|---|---|---|
| Social subsystem | Skills-Training, Governance & Ethics, Incentives, Cross-functional Teams | Tag at least one social code for every initiative. |

| Technical subsystem | Algorithm Type, Data-Infrastructure, HPC/Cloud, Cybersecurity | Tag at least one technical code for every initiative. |
|---|---|---|

Two trained coders independently tagged each document, and any disagreements were adjudicated by the third author. Coders also flagged the environmental context (Oil-Sector, Public-Sector Legacy, Expat Talent, Cultural/Language Needs) whenever it was explicitly mentioned. From these tags we calculated an "STS-alignment index", the share of initiatives within each country that carried both social and technical codes, providing a quantitative gauge of joint optimization. The STS-alignment index in the codebook is a mechanical ratio, measured in equation (2) as:

$$\text{STS-alignment Index} = \frac{Initiatives\ carrying\ both\ a\ Technical\ and\ a\ Social\ tag}{All\ initiatives\ in\ that\ country}$$

Inter-coder agreement reached Krippendorff's $\alpha = 0.82$, confirming the reliability of the coding frame (Krippendorff, 2016). We report proportions with 95% Wilson confidence intervals and provide numerators/denominators by country to indicate uncertainty, especially given the small sample sizes. All references are provided at the end of the codebook, and the complete document is available in the repository for replicability and transparency: Appendix A: 10.5281/zenodo.15802556.

*Exclusion criteria.* The document-harvesting phase returned 250 candidate items drawn from ministerial portals, stock-exchange filings, global news wires and corporate press rooms. We eliminated materials that (i) made no explicit reference to AI, machine learning or data-science capability; (ii) supplied insufficient detail to identify a discernible activity or outcome; (iii) lay outside the January 2017–April 2025 window; or (iv) were inaccessible in English or Arabic. Duplicate postings and password protected reports were likewise discarded. After this triage 47 records remained, forming the corpus for inductive coding. Two independent coders then tagged every entry with the study's three STS labels, Technical, Social, and Environment, and resolved disagreements in a reconciliation session.

*Illustrative coding samples.* Consider the 2019 press release announcing Bahrain's AI Academy at Bahrain Polytechnic:

> "*The AI Academy was launched in April 2019 to train Bahraini students in data analysis, data science and machine learning, thereby strengthening national AI capabilities and supporting the Kingdom's wider FinTech ecosystem*".

Under the STS protocol the initiative receives

- A social tag (Skills-Training) because the statement foregrounds human-capital development.
- A technical tag (Data science curriculum) since the program delivers instruction in data-science stacks.
- An environment tag (FinTech Ecosystem Driver) because the motive is to align with Bahrain's economic-diversification agenda.

This tri-fold tagging makes the activity comparable with, say, KSA's launch of the Noor LLM (Technical + Social, but weaker on Environment) or Oman's Vision 2040 education scholarships (Social + Environment, minimally Technical).

*The final documents breakdown.* To offer a clearer view of the 47 documents retained after exclusion, we present two summary tables below. Table 1 classifies these items by document type, such as (government strategy, press release, corporate announcement, or policy report), revealing how each category contributes to the overall dataset.

Table 2 organizes the same documents by country, indicating how many activities were captured in each of the six GCC states (KSA, UAE, Qatar, Kuwait, Bahrain, and Oman).

**Phase 3: Comparative case studies (micro work-system level).** To observe socio-technical design in practice, we compared the MBZUAI in the UAE and the SDAIA Academy in the KSA. Data were drawn from institutional reports, curricula, media coverage, and partnership announcements. Table 3 summarizes the five STS elements captured for each case. Each site was documented using a work-system template adapted from Cherns' (1976) STS principles.

**Table 1 Documents by Type.**

| Document Type | Count | % of Total |
|---|---|---|
| Government Strategy / Policy | 15 | 32% |
| Press Release | 17 | 36% |
| Corporate Announcement | 11 | 23% |
| Academic / Policy Report | 4 | 9% |
| Total | 47 | 100% |

**Table 2 Documents by Country.**

| Country | Number of Coded Activities |
|---|---|
| KSA | 10 |
| UAE | 11 |
| Qatar | 6 |
| Kuwait | 5 |
| Bahrain | 8 |
| Oman | 7 |
| Total | 47 |

**Table 3 STS Work-System Template for Case-Study Data Collection.**

| Element | Guiding STS question | Data collected |
|---|---|---|
| Tasks & workflows | How are tasks re-sequenced once AI tools are introduced? | Curriculum modules, project assignments, workflow charts. |
| People & skills | Which skills and autonomy levels are required at each step? | Admission criteria, graduate competencies, instructor profiles. |
| Technology | What AI models, data sources, and compute resources underpin the tasks? | LLM deployments, cloud/HPC specs, software stacks. |
| Governance & incentives | How are decisions, rewards, and accountability shared? | Funding model, certification pathways, partner MOUs. |
| Environment | How do oil-funded scholarships, expatriate policies, or Vision-2030 priorities shape the program? | Government directives, fiscal statements, labor-market targets. |

| Table 4 Operationalising the scenario axes through socio-technical indicators. | | |
|---|---|---|
| **Axis** | **Proxy indicators (data source)** | **STS subsystem primarily affected** |
| **Oil-revenue slack (horizontal)** | Brent crude annual average price; fiscal balance % GDP; sovereign-wealth-fund tech allocations (IMF, national MOFs) | **Technical** |
| **Regulatory harmonization (vertical)** | Presence of omnibus data-protection law; enforceable AI-ethics codes; cross-border talent-visa schemes; country-level STS-alignment index from Phase 2 | **Social** |

Cross case analysis examined joint-optimization quality (tight coupling vs. gaps) and equifinality (different socio-technical paths toward similar upskilling outcomes).

**Phase 4: Scenario analysis (dynamic system-evolution level)**. Building on evidence from Phases 1–3, this phase constructs a four-quadrant matrix that explores how GCC labor markets could evolve through 2030 under alternative combinations of oil-revenue slack and regulatory harmonization. In STS Theory terms, the first uncertainty governs the resource pool for the technical subsystem (HPC, data centers, frontier R&D), while the second reflects the maturity of the social subsystem (skills pipelines, incentive structures, governance routines). Joint optimization of these subsystems produces favorable AI outcomes; misalignment generates risk.

*Axes and operational indicators*. Table 4. lists the indicators that anchor each axis of the scenario matrix, oil-revenue slack for the technical subsystem and regulatory coherence for the social subsystem. These metrics act as heuristic signposts rather than forecasting inputs: they could position each GCC state within the four-quadrant framework to explore plausible socio-technical trajectories, but they are exploratory, not econometric, and should not be interpreted as point predictions.

**Limitations**. This study applies an STS lens to public-domain materials to assess AI-workforce preparedness across GCC states. Several constraints qualify our inferences. First, NAS's and related documents articulate official intentions but do not capture tacit routines within ministries or state-owned enterprises; we therefore analyze stated design premises, not internal practice. Second, the inductive corpus ($n = 47$) draws on publicly disclosed initiatives (2017–April 025) after triage from about 250 candidates, which introduces potential publication and survivorship bias (classified/defense or commercially sensitive projects are under-represented). Third, roughly 17% of NAS pages required OCR; although post-processing kept estimated transcription error under 5%, lexical artifacts could marginally affect TF–IDF salience and the subsequent STS mapping. Fourth, the STS alignment index measures whether an initiative carries both technical and social tags; it is a coverage indicator rather than a graded quality metric, and country-level estimates rest on small denominators (we therefore report numerators, denominators, and 95% Wilson CIs and avoid strict rank claims where intervals overlap). Fifth, the comparative case studies (MBZUAI; SDAIA Academy) illustrate archetypes (depth-first vs breadth-first) and are not intended to be exhaustive of GCC training models. Finally, the scenario matrix is heuristic, not a forecast; it abstracts from factors such as compute export controls, training-data availability/provenance, and unexpected regulatory or geopolitical shocks that could alter trajectories.

**Findings**
This section synthesizes the results from each methodological phase, policy document analysis, inductive codebook development, comparative case studies, and scenario planning through the STS lens. By triangulating quantitative and qualitative data sources, the findings underscore how GCC countries are preparing for AI-driven workforce transformations.

**Insights from the policy document analysis**
*Timeline of NAS adoption in the GCC*. Figure 8 charts more than publication dates; it chronicles the GCC's evolving socio-technical logic. The UAE's 2018 NAS launches the sequence with a hardware-first agenda, HPC, data-center build-out and pilot projects. Qatar's 2019 strategy retains that infrastructure focus but adds Arabic-focused and culturally specific applications, signaling that linguistic context is integral, not optional. KSA's 2020 NAS completes the early-adopter set by pairing large-scale infrastructure with the region's first quantified upskilling target for data- and AI-specialists, anchoring human-capital goals inside the technical rollout. Together these documents set the initial boundary conditions: capital-intensive hardware, data-sovereignty ambitions and a recognition that skills pipelines must keep pace.

Later entrants, Oman (2023), Bahrain (2024) and Kuwait (2024), inherit a well-defined technical lane and therefore pivot toward the social complement: ethics guidelines, SME-skilling grants, public-sector job redesign and talent-mobility schemes. In STS terms these are compensatory mechanisms to rebalance a system that began as technology-led.

The right-hand clustering of dates thus records collective learning. Early plans proved that sovereign wealth could deploy compute quickly; later plans acknowledge that, without parallel redesign of roles, incentives and governance, hardware alone fails to deliver labor-market returns. The region has shifted from a technology sprint to a socio-technical race in which durable success depends on synchronizing compute, data, skills and regulatory oversight within national visions such as Saudi Vision 2030 and UAE Centennial 2071 (Saudi Vision 2030, 2016; UAE Cabinet, 2023).

*The "Skill Genome"*. The "skill genome" analysis dissects each NAS into the two complementary strands that STS theory requires: a social strand that specifies people, roles, and rules, and a technical strand that specifies algorithms, data, and infrastructure. Using two successive TF-IDF passes, we first map the lexical density of social priorities and then of technical capabilities, thereby creating a paired snapshot of how far each document goes toward balancing its socio-technical design.

Social-subsystem priorities: In the first round of TF-IDF screening we isolated the social layer of each NAS. Three recurrent themes dominate: Skills Development, Job Creation, and Data Governance. Why these three? From an STS standpoint any AI strategy must explain (i) how people will learn (Skills Development), (ii) where those people will work once trained (Job Creation), and, because AI systems run on data, (iii) under what rules their data may circulate (Data Governance). Together they form the social counter-weight to the technical ambitions spelled out elsewhere in the documents. The heat-map in Fig. 9 quantify the lexical density of these themes, offering a rapid proxy for social-subsystem design across the six states.

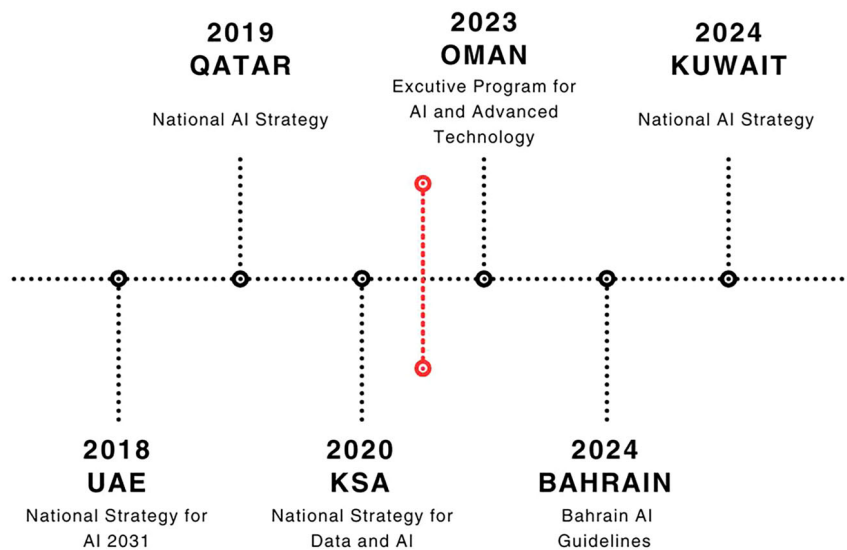

**Fig. 8 Timeline of National AI Strategy (NAS) adoption in the GCC.** Publication sequence (2018–2024) marking technical → socio-technical emphasis shift. Authors' original figure from NAS documents.

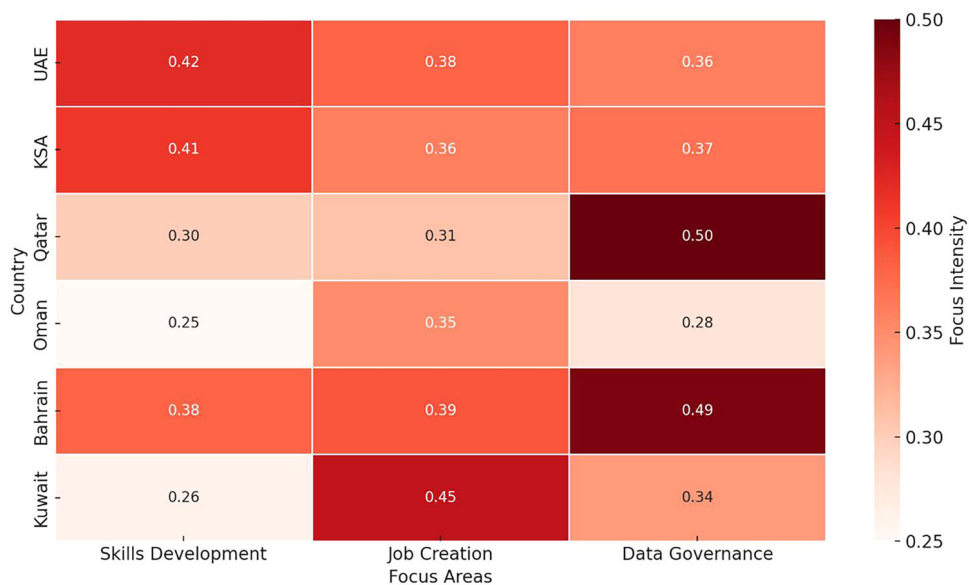

**Fig. 9 NAS 'social subsystem' focus (TF–IDF heat map).** Lexical salience of Skills Development, Job Creation, and Data Governance by country. Authors' analysis.

**Skills development**. All strategies endorse up- and re-skilling, yet TF-IDF scores reveal a gradient. The UAE (0.42) and KSA (0.41) occupy the top tier, reflecting explicit references to specialist graduate scholarships (UAE) and a sizeable national training commitment (KSA). The remaining four states cluster between 0.25 and 0.38, rich in rhetoric, sparer in concrete targets. In STS terms the region converges on a skills narrative, but only two members have translated that promise into measurable capacity-building programs (MBZUAI, SDAIA Academy).

**Job creation**. Scores are flatter but still differentiated. Kuwait peaks (0.45), consistent with its aim to shift citizens from the public to the private sector. Bahrain and Oman hover in the 0.35–0.39 range, offering aspirational language rather than task-level design. From an STS angle Kuwait's specificity signals tighter coupling between new technology and redesigned roles; elsewhere that linkage remains loose.

**Data governance**. A counter-intuitive pattern appears. The two states already enforcing omnibus personal-data laws, the UAE (2021) and KSA (2023), register only mid-range scores (around 0.36–0.37). Their strategies have moved beyond foundational governance toward operational roll-out, so the issue occupies fewer lines of text even as legal machinery is robust. Qatar (0.50) and Bahrain (0.49) top the chart precisely because they are still codifying privacy frameworks and devote more narrative space to the topic. The divergence illustrates an STS maxim: once a social control becomes institutionalized, its textual salience may decline even as practical enforcement strengthens. Nevertheless, data governance remains a fast-moving target, and because GCC governments themselves are the region's largest aggregators of citizen and service data, forthcoming regulation will need to sharpen oversight mechanisms for public-sector data controllers as much as for private-sector actors.

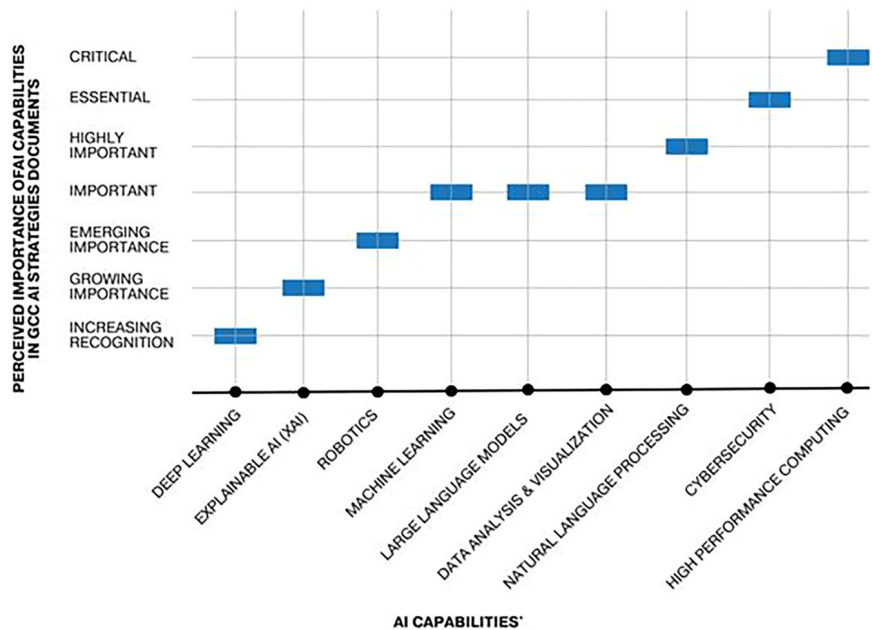

**Fig. 10 Technical AI capabilities in GCC AI strategies.** Ranked salience (HPC, cybersecurity, Arabic NLP, etc.) from TF–IDF screening. Authors' analysis.

| Table 5 Overview of AI-related Data Security Provisions in Existing GCC Laws. | |
| --- | --- |
| **Country** | **Law/Regulation** |
| UAE | UAE Federal Personal Data Protection Law (2021) Cybercrime Law (Federal Decree-Law No. 34 of 2021) |
| KSA | KSA Personal Data Protection Law (2023) |
| Oman | Royal Decree No. 6/2022 promulgating the Personal Data Protection Law (2022) |
| Qatar | Law No. 13 of 2016 on the Protection of Personal Data |
| Bahrain | Personal Data Protection Law (PDPL), Law No. 30 of 2019 |
| Kuwait | Data Privacy Protection Regulation No. 26 of 2024 |

Technical-subsystem priorities: The second TF-IDF screening shifted the focus to the technical subsystem, the counterpart of the social themes discussed earlier. Fig. 10 ranks nine capabilities by their textual salience, from "increasing recognition" to "critical". Read through an STS lens, three broad patterns emerge.

First, a strategic core. HPC, cyber security, and Arabic-oriented natural language processing occupy the top two bands of importance. Their prominence indicates that GCC governments view sovereign compute power, trusted data protection, and culturally appropriate language tools as preconditions for meaningful AI deployment. Yet every advance on the technical front immediately creates social corollaries: specialist operators, 24/7 monitoring teams, and compliance routines, thereby obliging the state to expand its skills base and governance capacity in parallel.

Second, an operational middle layer. Classic machine-learning techniques, LLMs, and data-analysis/visualization cluster in the "important" tier. These items suggest that most strategies have moved beyond pilot rhetoric toward mainstream adoption, yet their success will hinge on whether job frameworks and ML-Ops governance mature at comparable speed.

Third, an emergent periphery. Explainable AI, robotics, and deep learning register lower lexical weight. Their position does not imply irrelevance; rather, it signals that policy drafters either embed them within broader categories or consider them longer-term priorities, to be tackled once the strategic core is secured. This may also be because manufacturing as an industry is still developing in the region.

As with social themes, lexical salience is not a perfect proxy for implementation, but it does pinpoint where narrative energy, and likely funding, will flow.

*Overview of AI-related provisions in existing GCC Laws.* Table 5 summarizes the principal statutes that currently govern personal data, cyber-crime and, in a few instances, explicit AI practice across the six GCC members. In STS terms this corpus represents the formal control layer of the social subsystem: without such enforceable rules, any expansion of compute power or model complexity would rest on unstable ground.

Crucially, however, none of the GCC states has yet enacted an omnibus AI statute comparable to the EU AI Act adopted in 2024. The UAE Cabinet's decision in April 2025 to establish a comprehensive AI-linked legislative framework therefore marks meaningful progress, but remains an isolated initiative within the region.

A first reading of the table reveals a pronounced two-tier structure. At the upper tier sit the UAE and KSA. The UAE's 2021 Federal Personal Data Protection Law (PDPL), reinforced by Cyber-crime Law 34/2021, lays out lawful bases for processing, cross-border-transfer conditions and a schedule of cumulative fines that can escalate from AED 5000–5 million (approximately USD 1360–1.36 million) for missed API-registration deadlines or unreported breaches. KSA's 2023 PDPL mirrors that breadth, specifying data-subject rights, breach-notification windows and extra-territorial reach. Such provisions do more than deter misconduct; they create predictable behavioral routines, audit cycles, incident-response drills, vendor-assessment check-lists, that tie developers, data officers and compliance teams into a tightly coupled work system. In STS language, these two jurisdictions have begun to institutionalize the social controls needed to keep pace with their aggressive HPC roll-outs and domestic LLM initiatives.

The second tier, Kuwait, Qatar, Oman and Bahrain, relies on a patchwork of localization clauses, sector-specific cyber-security articles and guidance notes from specialized agencies (for example, Qatar's National Cyber Security Agency guidelines on AI adoption). Each instrument addresses a slice of the risk landscape, yet gaps remain at the interfaces where data, algorithms and human decision-makers intersect. The result is what socio-technical scholars describe as weak coupling: hardware and software can be purchased rapidly, but the liability assignments, escalation paths and incentive structures that give those technologies stable social meaning are left indeterminate.

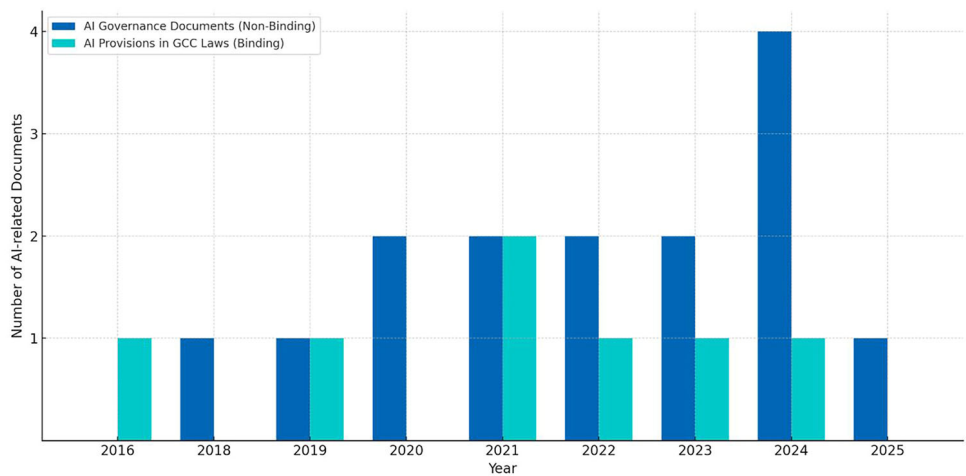

**Fig. 11 Growth in AI-related GCC documents (2016–2025).** Annual count of governance texts and statutory items. Authors' compilation.

Fragmentation matters because AI supply chains are intrinsically trans-national, cloud hosting, model fine-tuning and annotation services routinely cross borders. Where legal baselines diverge, firms face a costly dilemma: over-comply with the strictest regime, slowing time-to-market, or settle for the lowest common denominator, increasing residual risk. Either choice impedes the joint optimization that STS prescribes, leaving the technical subsystem under-utilized and the social subsystem over-stretched.

*Growth in AI-related documents in the GCC.* Figure 11 tracks two streams of documentary output from 2016 to 2025: (i) AI-governance documents, NASs updates, white papers, sectoral charters, voluntary ethics codes, and (ii) AI-specific provisions embedded in primary legislation. For most of the period the bars barely exceed one item per year; activity quickens in 2020 and then jumps in 2024, when four separate governance texts appear, while statutory change remains modest and irregular.

An STS reading suggests that this asymmetric growth is not accidental. Non-binding guidance allows ministries to broadcast desired norms, fairness, transparency, privacy, faster than the slower machinery of formal legislation. These texts widen the conversation but stop short of supplying the structural glue, clear liability chains, investigatory mandates, monetary penalties, needed to embed principles in day-to-day practice.

The gap shows up in two ways. First, many ethics codes demand "algorithmic transparency" yet omit audit procedures or escalation thresholds, leaving compliance teams with rhetorical aspirations but few operational tools. Second, overlapping guidance, health-care AI charters, telecom data-hosting notes, digital-authority white papers, can produce coordination noise. A start-up developing a medical-diagnosis model may struggle to know which document takes precedence, slowing deployment or prompting lowest-common-denominator compliance. In STS terms, the social subsystem becomes over-signaled but under-specified, making it hard to couple reliably with the fast-moving technical subsystem.

Even so, the surge in soft law has strategic value. Such instruments often operate as proto-statutes: they familiarize stakeholders with emerging obligations, shorten the learning curve when legislation eventually lands, and help cultivate the expert communities, lawyers, auditors, standards engineers, needed to enforce future rules. Whether the GCC's expanding stack of AI papers completes that transition will depend on the institutional capacity, outlined in "Overview of AI-related provisions in existing GCC laws", to convert exhortation into enforceable code. Until convergence occurs, the region's governance architecture remains symbolically rich but structurally light, leaving its rapidly growing technical base only partially stabilized by an equivalent social foundation.

**Emergent themes from the inductive content analysis**
*Codebook profile and analytic frame.* Appendix A consolidates 47 publicly disclosed initiatives launched between late 2017 and April 2025. Each entry is tri-tagged (Technical, Social, Environment) to gauge the degree of socio-technical joint optimization envisaged by its sponsors. Across the corpus, $34/47 = 0.72$ of initiatives (95% Wilson CI 0.58–0.83) carry both a Technical and a Social tag, our coverage proxy for balanced socio-technical design.

**Country-level indices (numerator/denominator; 95% CI) are as follows:**

- **KSA:** $9/10 = 0.90$ (0.60–0.98)
- **UAE:** $8/11 = 0.73$ (0.43–0.90)
- **Qatar:** $4/6 = 0.67$ (0.30–0.90)
- **Kuwait:** $3/5 = 0.60$ (0.23–0.88)
- **Bahrain:** $6/8 = 0.75$ (0.41–0.93)
- **Oman:** $4/7 = 0.57$ (0.25–0.84)

Two national patterns foreshadow the case-study results. KSA's 0.90 dual-tag share aligns with SDAIA's "compute-plus-talent" doctrine, while the UAE's 0.73 reflects several tech-heavy launches awaiting matching workforce scaffolding. The remaining states fall between 0.57 and 0.75, indicating uneven but real movement toward integrated design. Given small denominators, these differences should be interpreted as directional; intervals overlap and we avoid strict rank ordering.

*Note:* The alignment index indicates whether an initiative includes both social and technical elements. It does not grade the depth or quality of those elements; we therefore pair proportions with qualitative evidence below.

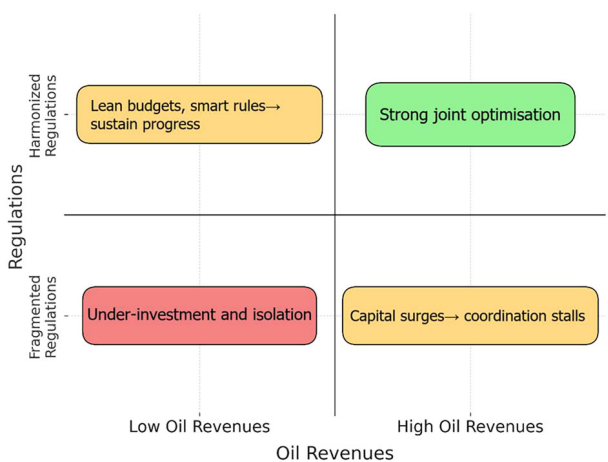

**Fig. 12 Scenario matrix for GCC AI socio-technical futures (2025–2030).**
2 × 2 matrix crossing oil-revenue slack and regulatory coherence; quadrant summaries. Authors' original figure.

*What the initiatives reveal about socio-technical evolution in the GCC.* **Workforce formation as first-order task.**

Well over half the initiatives foreground skills pipelines, specialist degrees at KAUST and MBZUAI; vocational boot camps at SDAIA; Microsoft-led workshops in Kuwait and the UAE; Bahrain's AI Academy for polytechnic students. From an STS perspective, these enlarge the human-capital pillar of the social subsystem and bind universities, sovereign funds, and vendors into cross-functional training networks. The pattern is consistent with the view that compute without competent operators will stall.

**A step-change in indigenous R&D capacity.**

The launches of Noor (KSA) and Jais (UAE) signal a pivot from "cloud consumer" to "foundation-model originator". Beyond HPC clusters and proprietary datasets (technical assets), both efforts created organizational units, SCAI in KSA; Inception and MBZUAI in Abu Dhabi, tasked with ethics review, benchmark testing, and Arabic-language curation. Here, the social and technical subsystems appear to advance in tandem, consistent with effective joint optimization.

**Sectoral diffusion and contextual tailoring.**

Finance, energy, health, and public administration dominate applied deployments, but each state bends AI toward its economic DNA. Bahrain leverages FinTech Bay's sandbox to pilot robo-advisers; Oman feeds predictive-maintenance models with oil-field telemetry; Qatar stress-tests cyber-defense AI ahead of mega-events; Dubai and NEOM embed conversational agents in municipal services. Such tailoring illustrates equifinality, different socio-technical paths toward comparable productivity goals, while also raising coordination challenges when common governance baselines are thin (see "Overview of AI-related provisions in existing GCC laws").

**Governance is catching up, but gaps persist.**

Only a minority of initiatives reference binding risk-management protocols or algorithm-impact assessments. A recurring note in the codebook is "guidelines under development", echoing the soft-law surge in Fig. 11. The implication is that while hardware budgets and LLM-branded headlines draw attention, governance routines remain the lagging component of the social subsystem. Unless that pillar hardens, the 0.72 alignment share risks plateauing. Recent steps, such as public moves in April 2025 to establish a comprehensive AI-linked legislative framework in the UAE, mark progress, but remain exceptions rather than the rule.

**Comparative case studies: research specialization Vs. vocational scale-up.** To examine socio-technical design at the institutional level, we compare two initiatives created in 2019 under strong sovereign sponsorship: the MBZUAI in the UAE and the SDAIA Academy administered by KSA's Data & AI Authority. Both pursue the goal of an AI-ready workforce, but they do so through markedly different configurations of technology, organization and labor-market interface, illustrating the range of pathways through which "joint optimization" can be attempted in practice.

*MBZUAI: research-driven depth.* MBZUAI is among the first graduate universities devoted exclusively to AI and has achieved top-100 placements in certain computer-science rankings as reported by the institution. A dedicated super-computing cluster anchors its technical subsystem, supporting peer-reviewed research, patent generation and, together with G42's Inception unit, the development of *Jais*, an open-source Arabic LLM. Organizationally, the university follows a partnership model: memoranda of understanding with global and regional actors, Google, Microsoft, IBM, Cerebras Systems, Presight AI, Abu Dhabi Global Market (ADGM), Silal, and, since 2025, École Polytechnique, direct externally defined problem statements and project funding into faculty laboratories. Outputs are validated in government sandboxes that test bilingual chatbots, public-health analytics, and supply-chain optimization, ensuring rapid transfer from discovery to application. Scholarships, a high faculty-to-student ratio, and an active visiting-scholar program reinforce the social subsystem, aligning incentives with advanced research.

*SDAIA Academy: vocational breadth.* KSA's National Strategy for Data & AI assigns SDAIA Academy the task of training 20,000 data and AI specialists by 2030. The Academy delivers modular courses, 2–12 weeks, in data analytics, cyber-security and applied machine learning through partnerships with Microsoft, IBM and leading KSA universities. Curricula are organized around concrete use-cases such as route optimization in logistics corridors, anomaly detection in industrial facilities and conversational agents for municipal services, and culminate in vendor-recognized certificates. Graduates move into public-sector entities and large commercial projects like NEOM. Employer surveys report strong job-readiness, although career progression often plateaus at practitioner level unless participants pursue further study or transition to R&D clusters managed by the SCAI. To address this ceiling the Academy has piloted advanced tracks in deep learning and Arabic NLP, although enrollments remain small relative to overall throughput.

*Socio-technical synthesis.* Taken together, MBZUAI and SDAIA Academy illustrate two distinct but complementary coupling logics within the GCC's emerging AI talent architecture. MBZUAI represents a depth-first design. Its dense compute stack, high faculty-to-student ratio and externally funded research contracts form a tightly integrated technical core. That core is immediately mirrored by a social subsystem of elite scholarships, joint laboratories and rapid policy test-beds, allowing discoveries to flow quickly into ministerial sandboxes and commercial spin-offs. The result is a socio-technical configuration optimized for scientific breakthroughs, exactly the kind of knowledge-production capacity the UAE needs to move from technology importer to technology originator.

SDAIA Academy, by contrast, embodies a breadth-first design. Modular courses, vendor-aligned curricula and a wide partner network translate KSA's strategy of training 20,000 data and AI specialists into rapid human-capital deployment. Its technical layer relies less on frontier GPU clusters and more on scalable cloud-based instructional platforms; its social layer prioritizes certification logistics and sectoral placement, feeding graduates directly into mega-projects such as NEOM and the Red Sea development. This yields a practice-centered pipeline that can be expanded quickly to meet Vision 2030's diversification timetable.

From an STS perspective each institution optimizes a different quadrant of the socio-technical space: MBZUAI emphasizes capability depth, SDAIA Academy emphasizes labor-market reach. Neither model is sufficient on its own, MBZUAI's selectivity restricts workforce volume, while SDAIA Academy's short-cycle format risks a skills ceiling, but together they outline a two-tier pipeline: research inventors upstream and applied integrators downstream. Building bridging mechanisms, credit transfers, joint fellowships, shared HPC access and coordinated career pathways, would allow learners and projects to move fluidly between tiers, preventing a bifurcated labor market in which advanced R&D and routine implementation evolve in isolation. Such bridges are the next frontier in achieving full socio-technical alignment of the GCC's AI workforce strategy.

**Scenario analysis**. Figure 12 cross-plots two macro uncertainties for 2025–2030. Oil-revenue slack defines how far governments can extend the technical subsystem, data centers, HPC clusters, research grants. Regulatory coherence captures the maturity of the *social* subsystem, skills pipelines, liability rules, talent-mobility schemes. However, the present analysis does not account for access to compute capacity, for example, the hardware deal the UAE concluded with the United States in 2025, nor does it assess the provenance or availability of training data. Although both the UAE and KSA possess logistical advantages that could facilitate large-scale data aggregation, neither government yet operates under fully transparent open-data policies or guidelines.

*High oil revenues + harmonized regulations.* With hydrocarbon prices firm and data-governance rules converging, sovereign-wealth funds channel surplus capital into research complexes such as MBZUAI and SCAI. A unified GCC tech-visa regime accelerates inbound talent flows, while mutually recognizable licenses let data and specialists circulate; vocational providers (SDAIA Academy) scale smoothly and labor mobility becomes frictionless. Technical and social subsystems reinforce one another, producing a self-sustaining innovation loop.

*High oil revenues + fragmented regulations.* Funding for HPC and LLM labs is abundant, but divergent legal baselines hinder audit harmonization, credential portability and cross-border data flows. Visa rules also splinter, so experts' clusters in UAE and KSA, while diffusion to smaller markets slows. Skill shortages persist in advanced fields because experts hesitate to relocate into uncertain regulatory environments, leaving the socio-technical system unbalanced.

*Low oil revenues + harmonized regulations.* Fiscal austerity trims mega-projects, yet governments pool resources and align standards. Shared visa schemes and joint training consortia cut duplication costs, allowing niche specialists to circulate even when public budgets are thin. The social subsystem, standards bodies, interoperability mandates, compensates for the slimmer technical layer.

*Low oil revenues + fragmented regulations.* Budget cuts delay data-center builds and scholarship schemes just as legal fragmentation raises transaction costs. Restrictive or inconsistent visas prompt reverse migration, public-sector retrenchment displaces mid-skill workers, and patchy upskilling programs fail to absorb them. R&D retreats to isolated enclaves, cross-border initiatives dry up and neither subsystem stabilizes the other, heightening the risk of social push-back against automation.

## Discussion
Evidence drawn from four empirical lenses shows the GCC inching away from an early technology-push posture toward the joint optimization that STS Theory treats as the hallmark of sustainable innovation. A region-wide alignment index of 72% signals tangible movement, yet the spread between KSA's 90% and Oman's 57% confirms that socio-technical equilibrium is still in flux. Crucially, the scenario matrix suggests that regulatory convergence plausibly governs systemic resilience more than capital abundance under our modeled conditions. In these scenarios, fragmented rules can offset even high oil revenues, whereas modest budgets combined with harmonized standards help sustain socio-technical loops.

Chronological analysis of NASs captures a clear discursive pivot. Early documents from the UAE (2018), Qatar (2019) and KSA (2020) lavish attention on data centers, supercomputing and pilot deployments; later strategies in Oman (2023), Bahrain (2024) and Kuwait (2024) devote proportionally more space to ethics codes, SME upskilling and public-sector job redesign. This migration aligns with the STS proposition that unchecked expansion of the technical subsystem eventually demands "requisite variety" in the social subsystem. Yet rhetorical convergence has not produced universal practice. Dual-tagged initiatives remain unevenly distributed, signaling that coupling work is still partial and path-dependent.

Project level findings illustrate equifinality, the STS idea that multiple socio-technical recipes can achieve comparable ends. Bahrain modernizes finance through regulatory sandboxes, Oman advances reservoir analytics, and Qatar secures sports mega-events via AI-enabled cyber-defense. Despite these varied trajectories, parity is missing. KSA attaches AI diffusion to a quantified reskilling goal of 20,000 practitioners, whereas Kuwait's forthcoming strategy lists aspirations without operational metrics. In such contexts, hardware assets risk underutilization as training, audit and mobility routines lag behind.

The comparison of MBZUAI and SDAIA Academy crystallizes two viable but incomplete coupling archetypes. MBZUAI's depth-first model allies dense compute with elite scholarships and industry research contracts, yielding frontier outputs such as the Jais LLM. SDAIA Academy's breadth-first model provides short-cycle, vendor-endorsed courses at scale, meeting immediate labor-market demand. Together they outline a two-tier pipeline, research inventors upstream and applied integrators downstream, but absent credit transfer, portable credentials and shared HPC pools, the workforce risks bifurcating into an R&D elite and a plateauing practitioner base.

Legal analysis spotlights a persistent control-layer lag. Only the UAE and KSA enforce GDPR-style boundary controls with fines and breach-notification windows; elsewhere, soft-law guidance proliferates faster than binding statute, an STS pathology described as "over-signal, under-specify". The scenario matrix confirms that fiscal slack cannot compensate for legal fragmentation, whereas rule convergence can buffer severe budget constraints by reducing coordination costs and facilitating cross-border talent mobility.

Situating six petro-states inside an STS frame produces three theoretical extensions. First, resource surpluses can obscure socio-technical deficits: abundant capital lets infrastructure outpace governance. Second, regulatory coherence operates as a multiplier of social-system capacity rather than a mere constraint on technological ambition. Third, equifinality is regionally bounded: heterogeneous rules create negative externalities that individual states cannot fully internalize, making collective coordination a structural variable in socio-technical alignment.

Four research avenues emerge. Longitudinal coupling studies should track how NAS commitments migrate, or fail to migrate, into operational routines, revealing subsystem lag and catch-up cycles. Ethnographic investigations inside ministries, university labs and vendor ecosystems could surface tacit coordination practices that document analysis misses, enriching our understanding of everyday socio-technical work-arounds. Comparative analyses with other resource-intensive regions, for instance, Norway's sovereign-wealth-fund governance or Alberta's energy-tech pivot, would test whether the GCC's hardware-lead, governance-lag pattern is intrinsic to petro-economies or more broadly generalizable. Finally, outcome-based metrics, tracking wage dispersion, internal mobility, algorithmic error rates, are needed to move the field from proxy indicators of alignment toward direct measures of socio-technical performance.

The study relies on publicly available material, omitting classified or proprietary initiatives, particularly in defense and energy. Dual-tag coding captures breadth rather than depth of coupling; the alignment index is therefore an early-warning gauge, not a definitive audit. Scenario narratives remain sensitive to volatile oil prices and exogenous regulatory shifts. These boundary conditions invite deeper fieldwork and triangulation.

In sum, GCC AI policy is leaving the era of isolated technology sprints and entering a more demanding socio-technical race. Whether the region achieves resilient, inclusive labor-market outcomes will depend less on the next tranche of GPUs and more on the institutional craftsmanship required to fuse technical ambition with robust social architecture.

## Conclusion
This study offered the first region-wide, STS-informed audit of how the six GCC states are preparing their workforces for an AI-intensive future. Combining lexical analysis of NASs, an inductively coded inventory of forty-seven initiatives, two contrasting talent case studies, and forward-looking scenarios, the research answers the guiding question: Are GCC governments matching rapid technical investment with an equally systematic build-out of the social subsystem?

The short answer is partly, but unevenly. A consolidated alignment index of 72% shows that most AI programs now embed at least basic upskilling and governance provisions alongside hardware roll-outs. Yet the spread, from KSA's 90% to Oman's 57%, reveals persistent gaps. Scenario analysis suggests that, under our modeled conditions, regulatory convergence plausibly binds outcomes more than fiscal capacity: without interoperable data-protection regimes and portable credentials, even oil-funded supercomputers can under-deliver, whereas modest budgets allied to coherent rules help sustain progress.

Two structural tensions therefore define the region's next chapter. First, a dual-track talent system is forming, research elites cultivated in high-depth nodes such as MBZUAI versus rapidly trained practitioners emerging from SDAIA-style academies. Without formal bridges, credit articulation, shared compute pools, transferable certifications, this bifurcation will harden. Second, outside the UAE and KSA the legal foundation remains a mosaic of voluntary codes, exposing large-scale technical deployments to reputational and operational risk.

For scholarship, the findings extend STS theory beyond its usual Western focus by showing that resource surpluses can mask socio-technical deficits and that rule coherence functions as a multiplier of social capacity. Methodologically, the work is bounded by its reliance on public-domain sources and breadth-oriented coding; deeper ethnographic and longitudinal investigations are required to verify whether current alignment scores translate into durable labor-market outcomes.

Ultimately, the GCC's AI build-out is an unfolding natural experiment in systemic coupling. Compute clusters, domestic LLMs and smart-city platforms are proliferating, but their public value will hinge on the parallel strengthening of skills pipelines, incentive structures and enforceable norms. The region has moved beyond isolated technology sprints; the task now is to finish the socio-technical race.

## From technology sprint to societal marathon: policy actions
The evidence presented across "Finding" and 5 confirms that the GCC's AI trajectory will generate durable public value only when technical ambition is matched by social, cultural, and environmental stewardship. On the strength of the region-wide STS-alignment score of 72 percent (Emergent themes from the inductive content analysis) and mindful of the "two-track" talent risk uncovered in the comparative case studies (Discussion), we propose five mutually reinforcing policy actions that convert rapid hardware gains into broader societal benefit.

First, national up- and reskilling programs should integrate AI capabilities into cultural-heritage and public-service curricula. Arabic-language NLP modules can be leveraged for heritage documentation, computer-vision techniques for artifact conservation, and data-analytics toolkits for service-delivery optimization in healthcare and urban planning. Linking human-capital formation to identity preservation and administrative responsiveness enlarges the social subsystem without compromising cultural integrity.

Second, the environmental footprint of the region's AI infrastructure must be curtailed through green procurement standards. Public subsidies for data centers and foundation-model projects should be contingent upon explicit efficiency metrics, most notably low power-usage-effectiveness thresholds and minimum renewable-energy shares. Such conditions would moderate escalating cooling loads in arid climates while aligning national strategies with COP-level sustainability pledges.

Third, a GCC-wide "AI Skills Passport" should be instituted to enable mutual recognition of certificates issued by flagship programs such as the SDAIA Academy and MBZUAI. Harmonized credentialing would reduce labor-mobility frictions, allow all six economies to marshal scarce specialist talent even under fiscal pressure, and underpin an integrated regional labor market.

Fourth, policymakers must bridge the nascent two-track talent system. Rotational fellowships that route SDAIA-trained practitioners through MBZUAI research laboratories and, conversely, embed MBZUAI doctoral candidates in high-throughput vocational settings, would foster cross-pollination. Coupled with shared HPC access and micro-credential credit transfer, these initiatives could transform the current bifurcation into a seamless, vertically integrated talent pipeline.

Finally, targeted micro-grants for small and medium-sized enterprises, start-ups, and civil-society organizations should catalyze bottom-up experimentation. Priority domains might include logistics, advanced manufacturing, and cultural preservation, ensuring that responsible AI adoption disseminates beyond mega-projects and elite institutions into the broader economic and social fabric.

Collectively, these measures re-cast the GCC's AI agenda from a rapid technology sprint into a long-distance socio-technical marathon, one that simultaneously safeguards cultural heritage, advances sustainability, and secures inclusive prosperity alongside productivity gains.

## Data availability

## Acknowledgements

This research received no specific grant from any funding agency, commercial or not for profit sectors.

## Author contributions

All authors contributed to the conception and design of the research. M.A conducted the data collection and analysis. M.A, M.S and O.A participated in the interpretation of the results and contributed to the preparation and revision of the manuscript. All authors have read and approved the final manuscript.

## Competing interests

The authors declare no competing interests.

## Ethical approval

This study analysed publicly available documents and aggregate program-level infor-
mation, for example, national AI strategies, policy papers, press releases, corporate
announcements, and media coverage, and conducted scenario modeling. It involved no
interaction with human participants, no collection or processing of identifiable personal
data, and no use of animals. Consequently, institutional ethical approval was not required
for this research.

## Informed consent

Not applicable. The research did not involve human participants or the collection of any
individual-level or identifiable personal data.

## Additional information

**Correspondence** and requests for materials should be addressed to
Mohammad Rashed Albous.

**Reprints and permission information** is available at http://www.nature.com/reprints